\documentclass{article}
\usepackage{latexsym}

\newcommand{\radio}{\mbox{\tiny radio}}
\newcommand{\sound}{\mbox{\tiny sound}}

\newcommand{\eg}{e.g.\ }
\newcommand{\ie}{i.e.\ }

\newtheorem{theorem}{Theorem}

\newenvironment{proof}{\noindent \textit{Proof.}}{\hfill $\Box$ \par \bigskip}

\begin{document}

\title{Impersonation with the Echo Protocol}
\author{Yoo Chul Chung \and Dongman Lee}
\maketitle

\begin{abstract}
  The Echo protocol tries to do secure location verification using
  physical limits imposed by the speeds of light and sound.  While the
  protocol is able to guarantee that a certain object is within a
  certain region, it cannot ensure the authenticity of further
  messages from the object without using cryptography.  This paper
  describes an impersonation attack against the protocol based on this
  weakness.  It also describes a couple of approaches which can be
  used to defend against the attack.
\end{abstract}

\section{Introduction}
\label{sec:intro}

Knowing the physical location of an entity can be very useful.  Such
knowledge is useful for location-based access control or context-aware
applications~\cite{bertino:sacmat2005,covington:sacmat2001,kaasinen:puc2003}.
However, we must be able to ensure that we have the correct location
of an entity for it to be a useful factor in access control.

Location determination is the problem of finding out where an entity
is located at.  In contrast, location verification verifies that an
entity is indeed located at where it claims to be, where the entity
somehow finds out the location by some other means.  If the location
determination mechanism is insecure, then we can use secure location
verification to ensure that an entity is located at a certain
location.

The Echo protocol~\cite{sastry:wise2003} is an inexpensive location
verification mechanism.  It is a packet-based distance bounding
protocol~\cite{brands:eurocrypt1993} which takes advantage of physical
limits imposed by the speeds of light and sound.  It is able to
guarantee that a certain entity is located within a certain region.

In this paper, we will show that the Echo protocol is vulnerable to
impersonation attacks.  We describe the Echo protocol in
section~\ref{sec:echo-protocol}.  In section~\ref{sec:impersonation}
we detail the vulnerability, while section~\ref{sec:defense} suggests
modifications to the protocol which help defend against impersonation
attacks.

\section{The Echo protocol}
\label{sec:echo-protocol}

In Sastry et~al.~\cite{sastry:wise2003}, which introduces the Echo
protocol, the entity wishing to prove its location claim is called the
\emph{prover}, while the entity which wishes to verify the claim is
called the \emph{verifier}.  The protocol verifies that the prover is
located within a region centered around the verifier, which is called
the verifier's \emph{region of acceptance}.  The protocol basically
works by measuring the round trip time of signals between the verifier
and the prover.  A round trip time that is too long would imply that
the prover is far away from the verifier.  A complete description is
shown in figure~\ref{fig:echo}.

\begin{figure}[bthp]
  \begin{tabbing}
    \hspace{0.2em} \= \hspace{0.8em} \= \kill
    \textsc{Communication Phase:} \\
    \> 1. \> \(p \stackrel{\radio}{\longrightarrow}
    \mbox{broadcast} : (l, \Delta_p)\). \\
    \>\> The prover broadcasts its claimed location $l$ \\
    \>\> and processing delay $\Delta_p$. \\
    \> 2. \> \(t_s \leftarrow \mathbf{time}()\). \\
    \>\> \(v \stackrel{\radio}{\longrightarrow} p : N\). \\
    \>\> A single verifier $v$ starts its timer and responds \\
    \>\> with a random nonce~$N$. \\
    \>\> We require \(l \in \textsl{ROA}(v, \Delta_p)\) and
    \(\Delta_p \geq \frac{n}{b_0}+\frac{n}{b_i}\). \\
    \>\> If no such verifier exists or $\Delta_p$ is invalid, \textbf{abort}. \\
    \> 3. \> \(p \stackrel{\sound}{\longrightarrow} v:N\). \\
    \>\> \(t_f \leftarrow \mathbf{time}()\). \\
    \>\> The prover echoes the nonce over ultrasound. \\
    \>\> The verifier records the finish time. \\
    \textsc{Verifier Computation Phase:} \\
    \> 4. \> \textbf{if} sent nonce differs from received nonce \\
    \>\> \hspace{1em} \textbf{return false} \\
    \> 5.\>\textbf{if} \(t_f - t_s > \frac{d(v,l)}{c}+
    \frac{d(v,l)}{s}+\Delta_p\) \\
    \>\> \hspace{1em} \textbf{return false} \\
    \> 6.\> Otherwise, \textbf{return true}
  \end{tabbing}
  \caption{Description of the Echo protocol.}
  \label{fig:echo}
\end{figure}

The prover initiates verification by broadcasting its location.  A
verifier whose region of acceptance includes the location is selected.
The verifier sends a nonce, which is a random bit string, to the
prover by radio.  After receiving the nonce, the prover sends it back
to the verifier by ultrasound.  If the nonce received by the prover is
not the same as the one it sent, or if it took too long for the reply
to be received, then the prover's location claim is rejected.

A reply is too late if it arrives later than what would be expected
from the signal travel time and processing delays at the prover.  The
time required for a radio signal to reach claimed location~$l$ from
verifier~$v$ is $\frac{d(v,l)}{c}$, where $c$ is the speed of light
and $d(v,l)$ is the distance between $v$ and $l$.  The time required
for a sound signal to reach $v$ from $l$ is $\frac{d(v,l)}{s}$, where
$s$ is the speed of sound.  With a maximum delay~$\Delta_p$ for the
prover to respond to a message it receives, it should take no longer
than $\frac{d(v,l)}{c}+\frac{d(v,l)}{s}+\Delta_p$ for the verifier to
receive a reply.

The prover should be unable to predict the nonce the verifier sends,
so it cannot send back a reply before the nonce is received from the
verifier.  Therefore, if the prover is not inside the verifier's
region of acceptance, it will take too much time for the reply to
arrive at the verifier.  This fact ensures that the prover must indeed
be within the region it claims to be if its location claim is verified
with the Echo protocol.

The Echo protocol itself does not require cryptography or time
synchronization between provers and verifiers.  It only needs a
reasonably precise clock in the verifier and the means to communicate
with radio and sound.  It also does not require prearranged setup
between verifiers and provers.  This makes the protocol suitable for
low-powered devices in spontaneous environments, \eg ubiquitous
computing or sensor
networks~\cite{kindberg:pervasive2002,hill:cacm2004}.

\section{Impersonation attack}
\label{sec:impersonation}

The fact that the Echo protocol does not require prearranged setup nor
cryptography is touted as an advantage.  However, the protocol in its
original form is unable to verify location claims in a useful way
under these conditions.  The problem is not that the protocol falsely
verifies a location claim for a prover, but rather that anything else
is unable to securely take advantage of the verification result.

An obvious way to use the Echo protocol for access control is to first
verify the location of the prover, and then to grant access to the
prover based on this verification result.  The specific steps are
outlined more concretely as follows, where we assume that the verifier
also handles access control:
\begin{enumerate}
\item The verifier confirms the location claim of the prover.
\item The prover sends an access request to the verifier.
\item The verifier grants access to the prover.
\end{enumerate}

One thing to note is that in the original paper, the prover's identity
is not included explicitly in the messages sent between the verifier
and the prover during location verification. However, we cannot simply
ignore the identity, since the verifier would later on have no way of
knowing whether an access request comes from the same object whose
location was verified.  Thus we will assume that the prover's identity
is included with all messages during location
verification.\footnote{The content and encoding of an identity are
  unimportant as long as the identity is unique to a prover and is the
  same during location verification and subsequence access requests.
  For example, it could be the frequency at which the prover
  communicates with the verifier.}  Incidentally, this also helps
provers and verifiers filter out messages that are not meant for them.

Unfortunately, we cannot hide the identity without resorting to
cryptography.\footnote{Even with cryptography, we must take care in
  how we use it.  For example, encrypting the identity with a
  non-random cipher would simply result in another form of identity.}
If we wish to avoid the expense of cryptography, then we must assume
that the identity can be exposed to an adversary.  The adversary can
use this identity to obtain illegal access in a location-based access
control system, using a form of run internal replay
attack~\cite{syverson:csfw1994}.  More concretely, the adversary can
impersonate the prover as follows:
\begin{enumerate}
\item The verifier confirms the location claim of the prover.  The
  adversary obtains the identity of the prover by eavesdropping on the
  communications between the verifier and the prover.
\item The adversary sends an access request to the verifier, with the
  identity of the prover attached to the request.
\item The verifier confirms that the entity who sent the access
  request is the prover, and mistakenly grants access to the
  adversary.
\end{enumerate}

The impersonation attack takes advantage of the fact that an adversary
can learn the identity of the prover from the messages exchanged
during location verification, which it can then use to create a valid
access request.  Theorem~\ref{thm:impersonate} expresses this
intuition formally.

\begin{theorem}
  \label{thm:impersonate}
  Let $i$ be the identity of the prover, $k_i$ a secret held by the
  prover, $F(i,k_i,m)$ a function which maps an arbitrary message (\eg
  an access request) to a format accepted by the verifier, and $m$ the
  actual message to the access control system.  An adversary can
  impersonate the prover when the following conditions hold:
  \begin{itemize}
  \item $i$ is exposed to the adversary during location verification.
  \item $F(i,k_i,m)$ is efficiently computable by anyone given only $i$
    and $m$.
  \item Any $F(i,k_i,m)$ sent to the verifier at any time is accepted
    as a valid message from the prover after location is successfully
    verified.
  \end{itemize}
\end{theorem}
\begin{proof}
  Assume that the prover with identity~$i$ has gone through location
  verification.  The adversary learns $i$, which he can use to compute
  $F(i,k_i,m)$ for any $m$.  Since location verification has already
  finished, when the adversary sends $F(i,k_i,m)$ to the verifier, it
  will be accepted as a valid message from the prover.  In other
  words, the adversary has successfully sent a message which the
  verifier believes to be from the prover.
\end{proof}

If we simply attach the identity along with the access request, we do
not need the secret~$k_i$, and $F$ in theorem~\ref{thm:impersonate}
can be expressed as $F(i,k_i,m)=(i,m)$, which can obviously be
efficiently computed by anyone who knows the identity~$i$ and a
message~$m$.  With such an $F$, the Echo protocol satisfies the
conditions in theorem~\ref{thm:impersonate}, so it is vulnerable to
impersonation.

If we want to prevent an adversary from forging a message, the prover
must use a secret known only to itself and perhaps the verifier.
Otherwise, the adversary would be able to efficiently compute
$F(i,k_i,m)$, since everything required by the prover for its
efficient computation would also be known to the adversary.

Also, the verifier must be able to retrieve the identity~$i$ and the
actual message~$m$ when it receives $F(i,k_i,m)$.  Without a previous
arrangement with the prover, the verifier will have to be able to do
this without knowing the secret~$k_i$.

If the verifier does not possess the secret~$k_i$, and $m$ cannot be
efficiently computed from only $F(i,k_i,m)$ alone, \ie without using
the identity~$i$, then we require public key cryptography, or
something at least as computationally expensive, as is implied by
theorem~\ref{thm:public-transform}.  Even if the verifier and the
prover prearrange to share the secret~$k_i$, we will still require
symmetric key cryptography or its equivalent, as is implied by
theorem~\ref{thm:secret-transform}.

\begin{theorem}
  \label{thm:public-transform}
  Let function~$F$ satisfy the following conditions:
  \begin{itemize}
  \item $F(i,k_i,m)$ can be computed efficiently given only $i$, $m$,
    and a secret $k_i$.
  \item $F(i,k_i,m)$ cannot be computed efficiently given only $i$ and $m$.
  \item $m$ can be computed efficiently given only $i$ and $F(i,k_i,m)$.
  \item $m$ cannot computed efficiently given only $F(i,k_i,m)$.
  \end{itemize}
  The function~$F$ can be used to create a public key cipher of
  equivalent computational cost and strength.
\end{theorem}
\begin{proof}
  Define $G(i,F(i,k_i,m))=m$.  By assumption, $G$ can be computed
  efficiently.  We can create a public key cipher~$P$ by defining the
  encryption function as $E((i,k_i),m)=F(i,k_i,m)$ and the decryption
  function as $D(i,c)=G(i,c)$, where $(i,k_i)$ is the private key and
  $i$ is the public key.  From the given conditions:
  \begin{itemize}
  \item Encryption can be done efficiently given private key $(i,k_i)$.
  \item Encryption cannot be done efficiently without private key
    $(i,k_i)$, since without $k_i$ we cannot construct $(i,k_i)$
  \item Decryption can be done efficiently given public key $i$.
  \item Decryption cannot be done efficiently without public key $i$.
  \end{itemize}

  Therefore $P$ is indeed a public key cipher.  It is easy to see that
  it would take the same amount of effort to break $P$ as $F$, and
  that the computational costs are equal.
\end{proof}

\begin{theorem}
  \label{thm:secret-transform}
  Let function~$F$ satisfy the following conditions:
  \begin{itemize}
  \item $F(i,k_i,m)$ can be computed efficiently given only $i$, $m$,
    and a secret $k_i$.
  \item $F(i,k_i,m)$ cannot be computed efficiently given only $i$ and $m$.
  \item $m$ can be computed efficiently given only $i$, $k_i$, and $F(i,k_i,m)$.
  \item $m$ cannot be computed efficiently given only $i$ and $F(i,k_i,m)$.
  \end{itemize}
  The function~$F$ can be used to create a symmetric key cipher of
  equivalent computational cost and strength.
\end{theorem}
\begin{proof}
  The proof is nearly identical to that of
  theorem~\ref{thm:public-transform}.

  Define $G(i,k_i,F(i,k_i,m))=m$.  By assumption, $G$ can be computed
  efficiently.  We can create a symmetric key cipher~$S$ by defining
  the encryption function as $E((i,k_i),m)=F(i,k_i,m)$ and the
  decryption function as $D((i,k_i),c)=G(i,k_i,c)$, where $(i,k_i)$ is
  the secret key.  From the given conditions:
  \begin{itemize}
  \item Encryption can be done efficiently given key $(i,k_i)$.
  \item Encryption cannot be done efficiently without key $(i,k_i)$,
    since without $k_i$ we cannot construct $(i,k_i)$.
  \item Decryption can be done efficiently given key $(i,k_i)$.
  \item Decryption cannot be done efficiently without key $(i,k_i)$,
    since without $k_i$ we cannot construct $(i,k_i)$.
  \end{itemize}

  Therefore $S$ is indeed a symmetric key cipher.  It is easy to see
  that it would take the same amount of effort to break $S$ as $F$,
  and that the computational costs are equal.
\end{proof}

If the message~$m$ can be easily retrieved from $F(i,k_i,m)$, the
other requirements for $F$ essentially require it to be a message
authentication code or a digital signature scheme, depending on
whether the secret~$k_i$ is shared or not, respectively.  Although
message authentication codes can be more efficiently than symmetric
cryptography~\cite{l:kaliski:cryptobytes1995}, digital signature
schemes typically require primitives from public key
cryptosystems~\cite{l:bellare:eurocrypt1996,l:NIST1994}

We can conclude that an adversary can impersonate the prover after
location verification is done using the Echo protocol, if we do not
allow the use of public key cryptography when there is no previous
setup between provers and verifiers.

\section{Defenses}
\label{sec:defense}

In this section we suggest a couple of approaches which can be used to
defeat impersonation attacks.  They are based on avoiding the
conditions listed in theorem~\ref{thm:impersonate}.  As in
\cite{sastry:wise2003}, we assume the verifier to be trustworthy.

\subsection{Cryptography}
\label{sec:crypto-defense}

The simplest way to prevent an impersonation attack is to just bear
the cost of cryptography.  Encrypting messages sent between the prover
and the verifier after location verification finishes ensures that an
adversary would not be able to send a valid forged message to the
verifier.

By pre-sharing a secret key between the prover and the verifier, they
can prevent adversaries from sending forged messages using symmetric
cryptography.  We do not even need to use the keyed variant of the
Echo protocol.\footnote{The keyed Echo protocol, which is described in
  the original paper, uses cryptography to guarantee that a
  \emph{specific} object is indeed where it claims to be, using a
  secret key pre-shared between the prover and identifier.}  The
verifier can use a table to look up the secret key associated with a
given identity.  An adversary would not be able to create a valid
forged message since it would not be able to get the secret key from
the identity.

We could also avoid the need for prearrangements between the prover
and verifier by using public key cryptography.  We could actually use
the \emph{public key} of the prover as the identity.  Using the Echo
protocol, the verifier can confirm that the public key belongs to a
prover that is located within its region of acceptance.  The adversary
cannot learn the private key, so it would not be able to send a forged
message that is seemingly sent by the prover.

One thing to note is that we must take care to ensure that encrypted
messages sent between the prover and the verifier cannot be used in a
replay attack.  This can be done using a challenge response protocol,
taking advantage of the secret or public key known by the verifier.
The design of such a protocol to resist replay
attacks~\cite{gong:csfw1993,aura:csfw1997,abadi:tse1996} is beyond the
scope of this paper.

\subsection{One-way Echo protocol}
\label{sec:prepend-defense}

While using cryptography is a simple solution to preventing
impersonation attacks, it is also an expensive one which negates one
of the important advantages of the Echo protocol, which is its frugal
use of resources.  Fortunately, we can modify the Echo protocol so
that it is resistant to impersonation attacks without having to use
cryptography.

The modification is very simple.  Instead of sending the message after
location verification, the prover sends the message when it first
initiates location verification.  By doing so, we circumvent the third
condition in theorem~\ref{thm:impersonate}, which requires that any
message sent after location verification finishes is accepted as
valid.  The modified protocol, which we will call the one-way variant
of the Echo protocol, is described in figure~\ref{fig:oneway-echo}.
We note that the identity of the prover which is implicitly included
in each message during location verification is no longer required for
location-based access control, although it is still useful for
filtering out irrelevant messages sent by unrelated provers and
verifiers.

\begin{figure}[bthp]
  \begin{tabbing}
    \hspace{0.2em} \= \hspace{0.8em} \= \hspace{1em} \= \kill
    \textsc{Communication Phase:} \\
    \> 1. \> \(p \stackrel{\radio}{\longrightarrow}
    \mbox{broadcast} : (m, l, \Delta_p)\). \\
    \>\> The prover broadcasts a message $m$, its claimed \\
    \>\> location $l$, and processing delay $\Delta_p$. \\
    \> 2. \> \(t_s \leftarrow \mathbf{time}()\). \\
    \>\> \(v \stackrel{\radio}{\longrightarrow} p : N\). \\
    \>\> A single verifier $v$ starts its timer and responds \\
    \>\> with a random nonce~$N$. \\
    \>\> We require \(l \in \textsl{ROA}(v, \Delta_p)\) and
    \(\Delta_p \geq \frac{n}{b_0}+\frac{n}{b_i}\). \\
    \>\> If no such verifier exists or $\Delta_p$ is invalid, \textbf{abort}. \\
    \> 3. \> \(p \stackrel{\sound}{\longrightarrow} v:N\). \\
    \>\> \(t_f \leftarrow \mathbf{time}()\). \\
    \>\> The prover echoes the nonce over ultrasound. \\
    \>\> The verifier records the finish time. \\
    \textsc{Verifier Computation Phase:} \\
    \> 4. \> \textbf{if} sent nonce differs from received nonce \\
    \>\>\> do nothing \\
    \> 5.\>\textbf{if} \(t_f - t_s > \frac{d(v,l)}{c}+
    \frac{d(v,l)}{s}+\Delta_p\) \\
    \>\>\> do nothing \\
    \> 6.\> Otherwise, process message $m$
  \end{tabbing}
  \caption{Description of the one-way Echo protocol.}
  \label{fig:oneway-echo}
\end{figure}

Since the message is sent immediately \emph{before} location
verification occurs, an adversary would not be able to send a forged
message that would be accepted by the verifier, simply because it
would not be able to predict when the prover would attempt location
verification.  Of course, the prover must avoid sending messages at
precisely predictable times, since an adversary could send a precisely
timed message with a very strong signal which ``overwrites'' the
message sent by the prover.

The one-way variant of the protocol has some limitations.  The most
important one is that location verification must be done for
\emph{every} message.  If there is a significant time interval between
the sending of a message and the start of location verification, an
adversary can send its own message during that time interval.  Without
using cryptography, the verifier would have no way of knowing that it
is not from the prover.  Since we would expect significant time gaps
between multiple messages, we would not be able to do location
verification once and expect the result to be valid for all of them.

Another limitation is that it is not useful for location-based access
control of information.  In order to retrieve information, the
verifier would have to send it to the prover.  If we send the
information in the clear, then the signal can be detected from outside
the verifier's region of acceptance, making access control pointless.
If we encrypt the information and send the ciphertext, then the
modification is unnecessary since we could have simply used the
original Echo protocol.

However, the one-way Echo protocol is useful for sending messages to
the verifier without requiring a reply.  Such messages are well suited
to specifying actions on physical objects at the location, \eg turning
on lights or lowering the volume of a speaker.  We only care that the
messages come from certain locations in such situations, and we are
not interested in concealing information, so there is no need for
cryptography.  The prover can send a single message one-way to the
verifier using the modified protocol, which is why we suggest calling
it the one-way Echo protocol.


Messages sent with the one-way Echo protocol need to be short.  If a
message is too long, an adversary will have enough time to notice the
transmission and overwrite later parts of the message with a strong
enough signal.  This limitation can be overcome by sending a command
to accept a message with a given hash during location verification.
The actual message would be sent separately, which the verifier would
check against the hash.

Despite these limitations, the one-way Echo protocol is an inexpensive
way to verify location claims, and is resistant to impersonation
attacks.  Like the original Echo protocol, the protocol itself does
not require cryptography nor prearranged setup between provers and
verifiers.  Unlike the original protocol, messages such as access
requests need not be encrypted, since message transmission is
integrated into the protocol itself without opening the protocol to
impersonation attacks.

\section{Conclusions}
\label{sec:conclude}

Location verification, which verifies location claims made by provers,
can be done using the Echo protocol.  The protocol is able to
guarantee that a prover whose location claim is verified is indeed
within the region it claims to be.  It is able to do this without
requiring expensive operations such as cryptography or time
synchronization, nor does it require previous arrangements between
provers and verifiers.

Unfortunately, we cannot securely take advantage of a verification
result without using cryptography.  We have shown that this is because
the Echo protocol has the following properties.  First, it does not
hide the identity of the prover.  Second, an adversary can forge a
message that can appear to be from the prover.  Finally, a valid
message can be received at any time by the verifier.  In fact, any
location verification protocol with these properties will be
vulnerable to impersonation attacks.

A simple way to defend against impersonation attacks is to lift the
restriction against using cryptography.  We suggest a one-way variant
of the Echo protocol as an alternative.  Although it is limited to
sending short messages to the verifier when no reply is expected, it
is resistant against impersonation attacks, while still maintaining
the low resource requirements and spontaneity of the original Echo
protocol.

\bibliography{strings,articles,proceedings,local}

\begin{thebibliography}{10}

\bibitem{abadi:tse1996}
Mart{\'\i}n Abadi and Roger Needham.
\newblock Prudent engineering practice for cryptographic protocols.
\newblock {\em IEEE Transactions on Software Engineering}, 22(1):6--15, January
  1996.

\bibitem{aura:csfw1997}
Tuomas Aura.
\newblock Strategies against replay attacks.
\newblock In {\em Proceedings of the 10th Computer Security Foundations
  Workshop}, pages 59--68. IEEE Computer Society Press, June 1997.

\bibitem{l:bellare:eurocrypt1996}
M.~Bellare and P.~Rogaway.
\newblock The exact security of digital signatures, how to sign with {RSA} and
  {R}abin.
\newblock In {\em Advances in Cryptology -- {EUROCRYPT} '96: Workshop on the
  Theory and Application of Cryptographic Techniques}, pages 399--416.
  Springer-Verlag, 1996.

\bibitem{bertino:sacmat2005}
Elisa Bertino, Barbara Catania, Maria~Luisa Damiani, and Paolo Perlasca.
\newblock {GEO-RBAC}: A spatially aware {RBAC}.
\newblock In {\em Proceedings of the Tenth {ACM} Symposium on Access Control
  Models and Technologies}, pages 29--37. ACM Press, 2005.

\bibitem{brands:eurocrypt1993}
Stefan Brands and David Chaum.
\newblock Distance-bounding protocols.
\newblock In Tor Helleseth, editor, {\em Advances in Cryptology -- {EUROCRYPT}
  '93: Workshop on the Theory and Application of Cryptographic Techniques},
  pages 344--359. Springer-Verlag, 1994.

\bibitem{covington:sacmat2001}
Michael~J. Covington, Wende Long, Srividhya Srinivasan, Anind~K. Dey, Mustaque
  Ahamad, and Gregory~D. Abowd.
\newblock Securing context-aware applications using environment roles.
\newblock In {\em Proceedings of the Sixth {ACM} Symposium on Access Control
  Models and Technologies}, pages 10--20. ACM Press, 2001.

\bibitem{gong:csfw1993}
Li~Gong.
\newblock Variations on the themes of message freshness and replay.
\newblock In {\em Proceedings of the 6th Computer Security Foundations
  Workshop}, pages 131--136. IEEE Computer Society Press, June 1993.

\bibitem{hill:cacm2004}
Jason Hill, Mike Horton, Ralph Kling, and Lakshman Krishnamurthy.
\newblock The platforms enabling wireless sensor networks.
\newblock {\em Communications of the ACM}, 47(6):41--46, June 2004.

\bibitem{kaasinen:puc2003}
Eija Kaasinen.
\newblock User needs for location-aware mobile services.
\newblock {\em Personal and Ubiquitous Computing}, 7(1):70--79, May 2003.

\bibitem{l:kaliski:cryptobytes1995}
B.~S. {Kaliski Jr.} and M.~J.~B. Robshaw.
\newblock Message authentication with {MD5}.
\newblock {\em CryptoBytes}, 1(1), 1995.

\bibitem{kindberg:pervasive2002}
Tim Kindberg and Armando Fox.
\newblock System software for ubiquitous computing.
\newblock {\em IEEE Pervasive Computing}, 1(1):70--81, January 2002.

\bibitem{l:NIST1994}
National~Institute of~Standards and Technology.
\newblock {FIPS} publication 186: Digital signature standard, 1994.

\bibitem{sastry:wise2003}
Naveen Sastry, Umesh Shankar, and David Wagner.
\newblock Secure verification of location claims.
\newblock In {\em Proceedings of the 2003 {ACM} Workshop on Wireless Security},
  pages 1--10. ACM Press, 2003.

\bibitem{syverson:csfw1994}
Paul Syverson.
\newblock A taxonomy of replay attacks.
\newblock In {\em Proceedings of the 7th Computer Security Foundations
  Workshop}, pages 187--191. IEEE Computer Society Press, June 1994.

\end{thebibliography}
\bibliographystyle{plain}

\end{document}